\documentstyle [11pt]{article}
\def\fnote#1#2{\begingroup\def\thefootnote{#1}\footnote{#2}\addtocounter
{footnote}{-1}\endgroup}

\begin{document}

\hfill{UTTG-09-13}

\hfill{TCC-0006-13}

\vspace{36pt}

\begin{center}
{\large {\bf {Goldstone Bosons as Fractional Cosmic Neutrinos}}}

\vspace{36pt}
Steven Weinberg\fnote{*}{Electronic address:
weinberg@physics.utexas.edu}\\
{\em Theory Group, Department of Physics, University of
Texas\\
Austin, TX, 78712}

\vspace{30pt}

\noindent
{\bf Abstract}
\end{center}

\noindent
It is suggested that Goldstone bosons  may be masquerading as fractional cosmic neutrinos, contributing about 0.39 to what is reported as the effective number of neutrino types in the era before recombination.  The broken symmetry associated with these Goldstone bosons is further speculated to be the conservation of the particles of dark matter.

\vfill

\pagebreak

The correlations of temperature fluctuations in the cosmic microwave background depend  on the effective number $N_{\rm eff}$ of neutrino species present in the era before recombination.  Although observations are certainly consistent with the expected  value $N_{\rm eff}=3$, there have been persistent hints in the data that the effective number may be somewhat greater.  WMAP9 together with ground-based observations (WMAP9 +eCMB)[1] gave $N_{\rm eff}= 3.89\pm 0.67$, while Planck together with the WMAP9 polarization data and ground-based observations (Planck+WP+highL)[2] gives $N_{\rm eff}= 3.36\pm 0.34$, both at the 68\% confidence level.  Is it possible that some nearly massless weakly interacting particle is masquerading as a fractional cosmic neutrino?

As a candidate for an imposter fractional neutrino, one naturally thinks of Goldstone bosons, associated with the spontaneous breakdown of some exact or nearly exact global continuous symmetry.  They would of course be massless or nearly 
massless, and the  characteristic derivative coupling of Goldstone bosons would make them weakly interacting at sufficiently low temperatures.  

Since Fermi statistics reduces the energy density of neutrinos relative to massless bosons by a factor 7/8, and $N_{\rm eff}$ lumps antineutrinos with neutrinos, a neutral Goldstone boson might look like $(1/2)/(7/8)=4/7$ of a neutrino.  But for this to be true, there is an important qualification: the Goldstone bosons must remain in thermal equilibrium with ordinary particles until after the era of muon annihilation, so that the temperature of the Goldstone bosons matches the neutrino temperature.  If Goldstone bosons went out of equilibrium  much earlier, then neutrinos but not Goldstone bosons would have been heated by the annihilation of the various species of particles of the Standard Model, and the contribution of Goldstone bosons to $N_{\rm eff}$ would be much less than 4/7.  As we shall see, there is a plausible intermediate possibility, that  the contribution of Goldstone bosons to $N_{\rm eff}$ would be $(4/7)(43/57)^{4/3}=0.39$.  To judge when the Goldstone bosons went out of thermal equilibrium, we need a specific theory[3].

We will consider the simplest possible broken continuous symmetry, a $U(1)$ symmetry associated with the conservation of some quantum number $W$.  All fields of the Standard Model are supposed to have $W=0$.  To allow in the simplest way for the breaking of this symmetry, we introduce a single complex  scalar field $\chi(x)$, neutral under $SU(3)\otimes 
SU(2) \otimes U(1)$, which carries a non-vanishing value of $W$.  With this field added to the Standard Model, the most general renormalizable Lagrangian is 
\begin{eqnarray}
{\cal L}& =& -\frac{1}{2}\partial_\mu \chi^\dagger\;\partial^\mu\chi+\frac{1}{2}\mu^2 \chi^\dagger\chi
-\frac{1}{4}\lambda \left(\chi^\dagger\chi\right)^2\nonumber\\&& 
-\frac{g}{4}\left(\chi^\dagger\chi\right)\left(\varphi^\dagger\varphi\right)+{\cal L}_{SM}\;,
\end{eqnarray}
where $\mu^2$,  $g$ , and $\lambda$ are real constants; ${\cal L}_{SM}$ is the usual Lagrangian of the Standard Model; and $\varphi=(\varphi^0,\varphi^-)$ is the Standard Model's scalar doublet.  
Experience with the linear $\sigma$-model shows that with a Lagrangian like (1), there are several diagrams in each order of perturbation theory that must be added up in order to give matrix elements that agree with theorems governing soft Goldstone bosons.  To avoid this, it is better to separate a massless Goldstone boson field $\alpha(x)$ and a massive ``radial'' field $r(x)$ by defining
\begin{equation}
\chi(x)=r(x)e^{2i\alpha(x)}\;,
\end{equation}
where $r(x)$ and $\alpha(x)$ are real, with the phase of $\chi(x)$ adjusted to make $\langle \alpha(x)\rangle=0$.  (The 2 in the exponent is for future convenience.)  The Lagrangian (1) then takes the form
\begin{eqnarray}
{\cal L}& =& -\frac{1}{2}\partial_\mu r\;\partial^\mu r+\frac{1}{2}\mu^2 r^2
-\frac{1}{4}\lambda r^4\nonumber\\&& 
-2r^2\partial_\mu\alpha\partial^\mu\alpha-\frac{g}{4}r^2\left(\varphi^\dagger\varphi\right)+{\cal L}_{SM}\;.
\end{eqnarray}

The $SU(2) \otimes U(1)$ symmetry of the Standard Model is of course broken by  a non-vanishing vacuum expectation value of the field $\varphi^0$, with a real zeroth-order value $\langle \varphi\rangle\simeq 247$ GeV.  
 The $U(1)$ symmetry of  $W$ conservation  is also broken if $(\mu^2-g\langle \varphi\rangle^2)/\lambda$ is positive, in which case $r$ gets a real vacuum expectation value,  given in zeroth order by
\begin{equation}
\langle r\rangle =\sqrt{m_r^2/2\lambda}\;,~~~~~m_r^2\equiv \mu^2-g\langle \varphi\rangle^2/2\;.
\end{equation}

In this formalism, the interaction of Goldstone bosons with the particles of the Standard Model arises entirely from a mixing of the radial boson with the Higgs boson.  There is  a  term $-g\langle\varphi\rangle\langle r\rangle\varphi'r'$ in the Lagrangian (3), where $r'\equiv r-\langle  r\rangle$ and $\varphi'\equiv {\rm Re}\varphi^0-\langle \varphi\rangle$ , so that the fields describing neutral spinless particles of definite non-zero mass are not precisely $\varphi'$ and $r'$, but
instead $\cos\theta\, \varphi'+\sin\theta \,r'$ and $-\sin\theta\,\varphi'+\cos\theta \,r'$, with the mixing angle given by
\begin{equation}
\tan\,2\theta=\frac{g\langle\varphi\rangle\langle r\rangle}{m_\varphi^2-m_r^2}\;.
\end{equation}
Since only one Higgs boson has been discovered at CERN[4], with what appear to be the production rate and decay properties expected in the Standard Model, this mixing must be weak.  We will make the assumption that $|
\tan2\theta|\ll 1$, and return soon to the question whether this is plausible.

This $\varphi-r$ mixing allows the Higgs boson to decay into a pair of Goldstone bosons.  The fourth term in (3) contains an interaction $(1/2\langle r\rangle)r'\partial_\mu\alpha'\partial^\mu\alpha'$, where $\alpha'\equiv 2\langle r\rangle\, \alpha$ is the canonically normalized Goldstone boson field.  Together with one vertex of the mixing term $-g\langle\varphi\rangle\langle r\rangle\varphi'r'$, this gives a partial width
\begin{equation}
\Gamma_{\varphi\rightarrow 2\alpha}=\frac{g^2\langle\varphi\rangle^2 m_\varphi^3}{16\pi (m_\varphi^2-m_r^2)^2} 
\end{equation}
Taking $\langle \varphi\rangle=247$ GeV, $m_\varphi=125$ GeV, and assuming $m_\varphi\gg m_r$, this partial width is $9.7\;g^2$ GeV.  The Goldstone bosons interact very weakly with particles of the Standard Model, so these decays would be unobserved.  But under the assumption that the production and  decays of the Higgs boson are correctly described by the Standard Model aside perhaps from decay into some new unobserved particles, the branching ratio for decay into new unobserved particles is known to be less than about 19\% [5], so with a Higgs width of about 4 MeV, the partial width (6) must be less than $0.8$ MeV, and therefore $|g|<0.009$.  With $g$ this small, and again assuming that $m_\varphi\gg m_r$, the mixing parameter (5) is indeed much less than one, provided that $\langle r\rangle$ is much less than 7 TeV, which seems not implausible.

Now, back to the problem of when the Goldstone bosons cease being in thermal equilibrium with the particles of the Standard Model.   
The joint action of the previously discussed terms  $-(1/2\langle r \rangle) r'\partial_\mu\alpha'\partial^\mu\alpha'$  and $-g\langle\varphi\rangle\langle r\rangle\varphi'r'$ in the Lagrangian (3)  produces an effective interaction between low-energy Goldstone bosons and any fermion ${\cal F}$ of the Standard Model:
\begin{equation}
-\frac{gm_{\cal F}}{2m_r^2m_\varphi^2}\partial_\mu\alpha' \;\partial^\mu\alpha'\;\overline{\cal F}{\cal F}\;.
\end{equation}
 At a temperature $T$, the derivatives in Eq.~(7) yield factors of order $kT$, and the number density of any particle with mass of order  $kT$ or less is of order $(kT)^3$, so the rate of collisions of Goldstone bosons with any species of fermion ${\cal F}$ with mass $m_{\cal F}$ at or below $kT$ is of order $g^2m^2_{\cal F}(kT)^7/m_r^4m_\varphi^4$.  The expansion rate of the universe is of order $(kT)^2/m_{\rm PL}$ where $m_{\rm PL}$ is the Planck mass, so the ratio of these two rates is
\begin{equation}
\frac{\rm collision}{\rm expansion}\approx \frac{g^2 m_{\cal F}^2(kT)^5m_{\rm PL}}{m_r^4 m_\varphi^4 }\;.
\end{equation}
This is a crude estimate, but the ratio decreases so rapidly with temperature that it gives a fair idea of when the Goldstone bosons go out of equilibrium.  

As mentioned earlier, if Goldstone bosons go out of equilibrium before $kT$  falls below the mass of most of the particles of the Standard Model, then the neutrinos (which are in thermal equilibrium at these temperatures) will 
be heated by the annihilation of Standard Model particles while the Goldstone bosons will not, and the contribution of Goldstone bosons to $N_{\em eff}$ will be much less than $4/7$.  But suppose that Goldstone bosons go out of equilibrium while $kT$ is still above the mass of muons and electrons but below the mass of all other particles of the Standard Model, a time when neutrinos are still in thermal equilibrium.  The cosmic entropy density just before the annihilation of muons, taking account of photons, muons, electrons, and three species of neutrinos, is
$$ s=\frac{4a_{\cal B}T^3}{3}\left(1+\frac{7}{4}+\frac{7}{4}+\frac{21}{8}\right)\;,$$
while after muon annihilation it is
$$ s=\frac{4a_{\cal B}T^3}{3}\left(1+\frac{7}{4}+\frac{21}{8}\right)\;,$$
where $a_{\cal B}$ is the radiation energy constant.
The constancy of the entropy per co-moving volume $s\,a^3$ tells us that for particles like neutrinos that are in thermal equilibrium, $Ta$ must increase by a factor $(57/43)^{1/3}$, while for free Goldstone bosons $Ta$ is constant, so that Goldstone bosons make a contribution to the measured $N_{\rm eff}$ equal to $(4/7)(43/57)^{4/3}=0.39$, which at least for the present seems in good agreement with observation.  For this to be the case, the ratio (8) must equal unity when $m_{\cal F}=m_\mu$ and $kT\approx m_\mu$, so that
\begin{equation}
\frac{g^2 m_\mu^7m_{\rm PL}}{m_r^4 m_\varphi^4 }\approx 1\;.
\end{equation}
For instance, with $g=0.005$ and $m_\varphi=125$ GeV, this tells us that $m_r\approx 500$ MeV.   (In order for  the Goldstone bosons to go out of equilibrium when the only massive Standard Model particles left are electrons and positrons, in which case they make a contribution to $N_{\rm eff}$ equal to $4/7$, the value of $m_r$ would have to be less than given by Eq.~(9) by a factor between  $(m_e/m_\mu)^{1/2}$ and $(m_e/m_\mu)^{7/4}$.)

Another consequence of the  term $-(1/2\langle r\rangle)r'\partial_\mu\alpha'\partial^\mu\alpha'$ in the Lagrangian is that the massive $ r$ bosons decay rapidly into Goldstone boson pairs.  Even for $\langle  r\rangle$ as large as 7 TeV, and taking $m_r=500$ MeV, the radial boson lifetime would be at most of order $10^{-16}$ seconds, so they would be long gone at any era with which we are concerned here.

We can further speculate about the physical significance of the assumed broken $U(1)$ symmetry.  There is no room for a new broken global symmetry in the Standard Model, so it natural to think of a symmetry associated with particles not described by the Standard Model, but known to be abundant  in the universe --- that is, with dark matter.  We will now assume that the conserved quantum number $W$ associated with the global $U(1)$ symmetry introduced above is WIMP number, the number of weakly interacting massive particles minus the number of their antiparticles.  We introduce a single complex Dirac WIMP field $\psi(x)$, carrying WIMP quantum number $W=+1$, and give the scalar field  $ \chi(x)$  WIMP quantum number $W=+2$, so that its expectation value leaves an unbroken reflection symmetry $\psi\rightarrow -\psi$.  All the fields of the Standard Model are again assumed to have $W=0$.  The most general renormalizable term involving the WIMP field that can be added to the Lagrangian (1) is
\begin{equation}
{\cal L}_\psi = -\bar{\psi}\gamma^\mu\partial_\mu\psi-m_\psi \bar{\psi}\psi -\frac{f}{2}\overline{\psi^c}\psi\,  \chi^\dagger-\frac{f^*}{2}\overline{\psi}\psi^c \, \chi\;,
\end{equation}
where $\psi^c$ is the charge-conjugate field\fnote{**}{That is, $\psi^c$ is the complex conjugate of $\psi$, multiplied by a matrix ${\cal C}^{-1}\beta$ (in the notation of ref. 6) that gives $\psi^c$ the same Lorentz transformation properties as $\psi$.};  $m_\psi$ and $f$ are constants; and by a choice of phase of $\psi$ we can make $f$ as well as $m_\psi$  real.
If together with the definition (2), we define a field $\psi'(x)$ by
\begin{equation}
\psi(x)=\psi'(x)e^{i\alpha(x)}\;,
\end{equation}
the WIMP Lagrangian (10) then becomes
\begin{eqnarray}
{\cal L}_\psi& =& -\overline{\psi'}\gamma^\mu\partial_\mu\psi'-m_\psi \overline{\psi'}\psi'-i\overline{\psi'}\gamma^\mu\psi'\;\partial_\mu\alpha\nonumber\\&& -\frac{f}{2}\overline{\psi'^c}\psi'\, r-\frac{f}{2}\overline{\psi'}\psi'^c \,r\;.
\end{eqnarray}
Because $r$ has a  non-zero vacuum expectation value $\langle  r\rangle$, the WIMP fields with definite mass are a pair of self-charge-conjugate fields
\begin{equation}
\psi_\pm(x)=\frac{1}{\sqrt{2}}\Big(\psi'(x)\pm \psi'^c(x)\Big)\;,
\end{equation}
with masses
\begin{equation}
m_\pm =m_\psi\pm \langle  r\rangle f\;.
\end{equation}
The part of the Lagrangian that involves the WIMP fields can then be put in the form
\begin{eqnarray}
{\cal L}_\psi& =& -\frac{1}{2}\sum_\pm\left[\overline{\psi_\pm}\gamma^\mu\partial_\mu\psi_\pm+m_\pm \overline{\psi_\pm}\psi_\pm\right]-\frac{i}{2}\left[\overline{\psi_+}\gamma^\mu\psi_-+\overline{\psi_-}\gamma^\mu\psi_+\right]\;\partial_\mu\alpha\nonumber\\
&&
-\frac{f}{2}r'\left[\overline{\psi_+}\psi_++\overline{\psi_-}\psi_-\right]\;,
\end{eqnarray}
where again, $
r'\equiv r-\langle r\rangle
$.

We see that instead of one Dirac WIMP, there are two Majorana WIMPs of different mass.  But the heavier WIMP will decay into the lighter one by emitting a Goldstone boson, while the lighter one  is kept stable by an unbroken reflection symmetry, so in this theory we can expect that the present universe will contain only one kind of Majorana WIMP,  the lighter one $w$, 
with mass $m_w$ equal to the smaller of $m_\pm$.

The $r-\varphi$ mixing allows the Higgs boson to decay into pairs of the lighter WIMPs, if they are lighter than $m_\varphi/2$.  In this case, the partial width for this decay is
\begin{equation}
\Gamma_{\varphi\rightarrow 2w}=\frac{1}{32\pi}\left(\frac{fg\langle r\rangle\langle\varphi\rangle}{m_\varphi^2-m_r^2}\right)^2\sqrt{m_\varphi^2-4m_w^2}
\end{equation}
As we have seen, observations require this to be less than about $0.8$ MeV.  Taking $ m_r$ and $2m_w$ much less than $m_\varphi$, this condition tells us that the WIMP mass splitting $\Delta m\equiv |m_+-m_-| =2|\langle r\rangle f|$ satisfies $ |g|\Delta m < 3.2$ GeV, a constraint that will be useful in what follows.

The surviving WIMPs can annihilate in pairs through their interaction with Goldstone bosons and with the field $ r'$, which mediates  interactions both  with Goldstone and radial bosons and with the particles of the Standard Model.  
It is well known that in order for annihilation of WIMPs to give a dark matter density like that observed, it is necessary for the annihilation cross-section to satisfy[7] 
\begin{equation}
m_w\left(\frac{2\pi\sum\langle\sigma v\rangle}{G_{\rm wk}^2 m_w^2}\right)^{0.51}\simeq 3.7\,{\rm GeV}\times (2\Omega_Dh^2)^{-0.54}\simeq 9\;{\rm GeV}\;,
\end{equation}
where   $\Omega_Dh^2\simeq 0.105$ is the usual dark matter density parameter; the sum is taken over all annihilation channels; and $G_{\rm wk}\simeq 10^{-5}\; {\rm GeV}^{-2}$  is the weak coupling constant.  In what follows we will simplify our estimates by replacing the exponent $0.51$ with $1/2$.

One possibility is annihilation into a quark $q$ and its antiparticle.    The combination of the interactions $f\overline{\psi_\pm}\psi_\pm r'$, the mixing term $-g\langle\varphi\rangle\langle r\rangle\varphi'r'$ and the Standard Model interaction $(m_q/\langle \varphi\rangle)\overline{q}q\varphi'$ gives an effective cross section for  annihilation of cold WIMP pairs into a relativistic quark $q$ and its antiquark:
\begin{equation}
\sum\langle \sigma v\rangle= \frac{3}{2\pi}\left(\frac{g m_q m_w\,\Delta m}{2(4m_{w}^2-m_r^2)(4m_{w}^2-m_\varphi^2)}\right)^2\;,
\end{equation}
in which we have used Eq.~(14) to express $|\langle r\rangle f|$ as $\Delta m/2$.

 For heavy WIMPs, with $m_w$ much larger than the mass $m_t$ of the top quark,  the quark produced in WIMP annihilation would  be  the top quark, in which case Eq.~(17) (with $2m_w$ much larger than $m_r$ and $m_\varphi$) requires that $\Delta m/m_w=32 m_w^2 G_{\rm wk}/\sqrt{3}|g|m_t\times 9\,{\rm GeV}\gg 32$, requiring  $|m_\psi|$ and $|\langle r\rangle f|$ to differ by much less than 6\%.  

 The fine tuning problem is worse  for $m_t>m_w\gg m_\varphi/2$.  In this case the quark produced in WIMP annihilation would be the bottom quark, and  Eq.~(17)  requires that 
$\Delta m/m_w =32 m_w^2 G_{\rm wk}/\sqrt{3}|g|m_b\times 9\,{\rm GeV}\gg 160$, which would require   $|m_\psi|$ and $|\langle r\rangle f|$ to differ by much less  than $1\%$.

 The case  $m_\varphi\gg 2m_w\gg m_r$ is even less promising.  In this case,  Eq.~(17) gives
$ m_w\simeq \sqrt{3}m_q\,|g|\,\Delta m/8m_\varphi^2\,G_{\rm wk}(9\;{\rm GeV})\;.$
With the previously derived upper bound  $|g|\Delta m<3.2$ GeV,  this  requires that $m_w<0.49 \,m_q$, which is clearly impossible if cold $w$ pairs are to annihilate into $q+\overline{q}$.

It appears that if $\Delta m$ and $m_w$ are of comparable magnitude, then the annihilation of these WIMPs into quarks may not be sufficiently fast to bring the dark matter density down to the observed value.  Inclusion of  annihilation into leptons and gauge bosons helps this problem, but apparently not enough.  Of course, we could  make the annihilation cross-section as large as we like by taking $2m_w$ sufficiently close to $m_\varphi$ (or $m_r$).  Otherwise,  the dominant annihilation could be into pairs of Goldstone bosons (and perhaps radial bosons).    The cross-section here is of order $f^4/m_w^2$, so condition (17) would require that $m_w\approx  10^4\;f^2$ GeV.  

Unfortunately there are too many free parameters here to allow a definite conclusion whether the density of WIMPs in this theory does or does not match the observed density of dark matter.

\vspace{20pt}

I am grateful for a helpful correspondence with Eiichiro Komatsu, for a valuable suggestion by Jacques Distler, and for information provided by Can Kilic and Matthew McCullough.  This material is based upon work supported by the National Science Foundation under Grant Number PHY-0969020 and with support from The Robert A. Welch Foundation, Grant No. F-0014.

\begin{center}
{\bf ----------}
\end{center}

\vspace{10pt}

\begin{enumerate}
\item G. Hinshaw {\it et al.} [WMAP collaboration], arXiv: 1212.5226; S. Das {\it et al.} [ACT collaboration], Astrophys. J. {\bf 729}, 62 (2011); R. Keisler {\it et al.} [SPT collaboration], Astrophys. J. {\bf 743}, 28 (2011).
\item  P. A. R. Ade {\it et al.} [Planck collaboration], arXiv:1303.5076; S. Das {\it et al.} [ACT collaboration], arXiv:1301.1037; C. L. Reichardt {\it et al.} [SPT collaboration], Astrophys. J. {\bf 755}, 70 (2012).  
\item The possibility that axion-like Goldstone bosons contribute to $N_{\rm eff}$ was mentioned along with other possible contributions by K. Nakayama, F. Takahashi, and T. T. Yanagida, Phys. Lett. B {\bf 697}, 275 (2011), without addressing the question of thermal equilibrium between these Goldstone bosons and Standard Model particles.
\item A. Aad {\em et al.} [ATLAS Collaboration], Phys. Lett. B {\bf 716}, 1 (2012); S. Chatrchyan {\em et al.} [CMS Collaboration], Phys. Lett. B {\bf 716}, 30 (2012).
\item P. P. Giardano, K. Kannike, I. Masina, M. Raidal, and A. Strumlo, arXiv:1302.3570.
\item S. Weinberg, {\em The Quantum Theory of Fields - Volume I} (Cambridge University Press, Cambridge, UK, 1995), Sec. 5.4.
\item B. W. Lee and S. Weinberg,  Phys. Rev. Lett. {\bf 39}, 165 (1977); D. D. Dicus, E. W. Kolb, and V. L. Teplitz, Phys. Rev. Lett. {\bf 39}, 168 (1977); E. W. Kolb and K. A. Olive, Phys. Rev. D {\bf 33}, 1202 (1986).  Eq.~(17) is adapted from Eq.~(3.4.14) of S. Weinberg, {\em Cosmology} (Oxford University Press, Oxford, 2008).  An additional factor 2 has been inserted in front of $\Omega_Dh^2$, because in this theory the present dark matter density consists of a  single Majorana WIMP species, rather than distinct particles and antiparticles as assumed in the derivation of Eq.~(3.4.14).
\end{enumerate}

\end{document}